\documentclass[lettersize,journal]{IEEEtran}
\usepackage{amsmath,amsfonts}
\usepackage{algorithmic}
\usepackage{algorithm}
\usepackage{array}
\usepackage[caption=false,font=normalsize,labelfont=sf,textfont=sf]{subfig}
\usepackage{textcomp}
\usepackage{stfloats}
\usepackage{url}
\usepackage{verbatim}
\usepackage{graphicx}
\usepackage{cite}
\usepackage{orcidlink}
\usepackage{hyperref}
\graphicspath{{./}{figures/}}

\newcommand{\nraoSMD}{Spectrum Management Department}
\newcommand{\nraoCV}{National Radio Astronomy Observatory (NRAO), Charlottesville, VA, USA}
\newcommand{\nraoNM}{NRAO Domenici Socorro Operations Center (DSOC), Socorro, NM, USA}
\newcommand{\gbo}{Green Bank Observatory (GBO), Green Bank, WV, USA}
\newcommand{\spaceX}{Space Exploration Technologies Corporation (SpaceX), Hawthorne, CA, USA}

\newcommand{\bnhan}{Bang D. Nhan}
\newcommand{\cdepree}{Christopher G. De Pree}
\newcommand{\abeasley}{Anthony Beasley}
\newcommand{\mwhitehead}{Mark Whitehead}
\newcommand{\kryan}{Kevin Ryan}
\newcommand{\dfaes}{Daniel Faes}
\newcommand{\tchamberlin}{Thomas Chamberlin}
\newcommand{\dpattison}{Dawn Pattison}
\newcommand{\vcatlett}{Victoria Catlett}
\newcommand{\alawson}{Aaron Lawson}
\newcommand{\dbautista}{Daniel Bautista}
\newcommand{\swasik}{Sheldon Wasik}
\newcommand{\fschinzel}{Frank Schinzel}
\newcommand{\ddueri}{Daniel Dueri}
\newcommand{\miverson}{Matt Iverson}
\newcommand{\jdonenfeld}{Jacob Donenfeld}
\newcommand{\bschepis}{Brian Schepis}
\newcommand{\dgoldstein}{David Goldstein}
\newcommand{\csteele}{Christopher Steele}

\def\gridline #1{\hspace*{0.05\hsize} \hbox to \hsize {#1} }

\begin{document}

\title{Lessons learned from ongoing coordination between \\ NRAO/GBO and LEO NGSO satellite constellations}

\author{
    \IEEEauthorblockN{
    \bnhan\orcidlink{0000-0001-5122-9997}\IEEEauthorrefmark{1}, 
    \cdepree\orcidlink{0000-0002-0786-7307}\IEEEauthorrefmark{1},
    \swasik\orcidlink{0000-0001-9213-0117}\IEEEauthorrefmark{1}, \\
    \alawson\orcidlink{0009-0000-0879-5125}\IEEEauthorrefmark{1},
    \dbautista\orcidlink{0009-0007-3897-2912}\IEEEauthorrefmark{1}$^{,}$\IEEEauthorrefmark{2},
    \fschinzel\orcidlink{0000-0001-6672-128X}\IEEEauthorrefmark{1}$^{,}$\IEEEauthorrefmark{3}}

    \IEEEauthorblockA{\IEEEauthorrefmark{1}\nraoSMD,~\nraoCV}
    
    \IEEEauthorblockA{\IEEEauthorrefmark{2}\gbo}

    \IEEEauthorblockA{\IEEEauthorrefmark{3}\nraoNM}
}

\maketitle

\begin{abstract}
With the increased demand for spectrum usage in recent years, particularly with the expansion of NGSO satellite systems for broadband internet and cellular services, radio astronomy observatories located at remote sites have been faced with the challenge of increasingly detrimental RFI exposure from ubiquitous satellite downlinks. With the help of the U.S. NSF, NRAO and the radio astronomy community have spearheaded collaboration efforts with the commercial NGSO operators to develop different coordination techniques to alleviate the impacts of these downlink signals in radio astronomy scientific observations. This paper summarizes the ongoing coordination efforts, spectrum interference avoidance schemes, and a case study on community participation of the Operational Data Sharing (ODS) system between RAS and NGSO operators. We highlight some of the essential lessons learned from these collaborations and potential application of the ODS system to other potential spectrum stakeholders.
\end{abstract}


\section{Introduction}
\label{sec:intro}
Within the past seven years, the boom in satellite-based broadband internet and cellular service has resulted in an exponential increase in the number of low-earth orbit (LEO, altitude from 160--2000~km), and non-geostationary orbit (NGSO) satellites (from 120 in December 2019 to $>$10,000 in March 2026)\footnote{Figures queried from \url{https://satellitemap.space/}.}. This number is expected to increase by many orders of magnitude in the upcoming decade \cite{walker2020satcon1, boley2021satellite, falle2023one}. This large increase has posed significant challenges to both optical and radio astronomy observatories (RAOs) on the ground \cite{mcdowell2020leo} and in space \cite{borlaff2025satellite}. For example, ubiquitous NGSO downlink (DL) signals have reduced the ``radio-quietness'' around ground-based radio telescope sites, which traditionally rely on geographical remoteness from large populations to reduce the amount of unwanted human-made radio frequency interference (RFI) for their scientific measurements. This impact will continuously be assessed by NRAO and other RAOs as more satellite constellations being launched.

In the past several years, as part of the National Science Foundation (NSF)’s Spectrum Innovation Initiative: National Radio Dynamic Zones (SII-NRDZ) \cite{nsf2022siinrdz} and Spectrum and Wireless Innovation enabled by Future Technologies - Satellite-Terrestrial Coexistence (SWIFT-SAT) \cite{nsf2023swiftsat}, through coordination agreements with NSF, NRAO has led the development of coordination techniques in conjunction with SpaceX and other NGSO satellite operators to achieve dynamic spectrum sharing and spectrum coexistence. 

In the U.S., these coordination efforts were originally part of the requirements set by the FCC Rule Footnote US131 \cite{fccus131}: ``\textit{The band 10.7-11.7~GHz, non-geostationary satellite orbit licensees in the fixed-satellite service (space-to-Earth), prior to commencing operations, shall coordinate with the following radio astronomy observatories to achieve a mutually acceptable agreement regarding the protection of the radio telescope facilities operating in the band 10.6-10.7~GHz.}'' The NSF/NRAO/NGSO coordination has recently expanded to the Supplemental Coverage from Space (SCS) service bands (a.k.a, Direct-to-Cell/DTC or Direct-to-Device; e.g., 1990-1995~MHz for Starlink/T-Mobile) \cite{farrar2025spectrum} and other frequency ranges observed by NRAO radio telescopes in hopes of meeting the protection criteria used for radio astronomical measurements set forth in ITU-R RA.769-2 \cite{itur769}. 

In this paper, we present an overview of the coordination processes in Section~\ref{sec:coordination}, and avoidance schemes in Section~\ref{sec:avoidance}. We focus on the community adoption of the Operational Data Sharing (ODS) framework as a case study of coordination in Section~\ref{sec:case_study}. Subsequently, we provide some general lessons learned from our coordination experience with NGSO and other stakeholders in Section~\ref{sec:lessons} before closing with outlook on future paths and coordination possibilities with other active spectrum users, including terrestrial transmitters in Section~\ref{sec:outlooks}.

\section{Coordination processes}
\label{sec:coordination}
Coordination efforts between US-based radio astronomy service (RAS) and NGSO operators have been led by the NSF. NSF has signed coordination agreements with NGSO operators, including SpaceX \cite{nsf2023spacex}, Amazon \cite{nsf2025amazon}, and AST SpaceMobile \cite{nsf2025ast}, for NSF astronomy facilities including the NRAO's Karl G. Jansky Very Large Array (VLA, an array of 27$\times$25-m dishes central New Mexico), Robert C. Byrd  Green Bank Telescope (GBT, a single 105~m diameter dish in West Virginia), and Very Long Baseline Array (VLBA, an array of 10$\times$25-m dish stations distributed from Mauna Kea, Hawai`i to St. Croix, U.S. Virgin Islands). 
%Changed to 27 - we usually do not include the spare in the headcount

Additionally, NSF has helped to facilitate coordination between other US-based radio observatories and the NGSO operators, including the SETI Hat Creek Radio Observatory (HCRO), MIT Haystack Observatory, Caltech's Owens Valley Radio Observatory (OVRO/DSA-2000), and K\={o}ke`e Park Geophysical Observatory (KPGO). Each facility's home institution and relevant parties have subsequently signed coordination agreements and, if applicable, non-disclosure agreements (NDA) with the NGSO operators separately. 

In practice, RAOs' engagement with the NGSO operators can often start well before the deployment of a satellite constellation. This provides adequate time for review of the technical information, and for mitigation techniques to be agreed upon as a coordination agreement is being formulated. At this stage in the coordination process, it is understood between RAOs and the NGSO operator that the theoretical results are subject to change after real-world testing. 

Once a coordination agreement, and/or NDA, is in place, each radio observatory can agree on different types of testing scenarios with the NGSO operators to conduct a suite of experiments to assess the impact of the downlink (DL) signals on their respective telescopes. If possible, spectrum avoidance schemes and operational modes may be preliminarily agreed upon and explored with follow-up tests. Once adopted, continuous or periodic monitoring of the radio telescope science data quality in the presence of the NGSO systems can provide actionable information on the effectiveness in reducing RFI resulting from the coordination. This monitoring can also provide an early warning for impacts of any changes resulting from newly launched satellite systems (both hardware and software). 

Thus far, NRAO has conducted two primary types of experiments with NGSO operators: coordinated and uncoordinated. The former involves careful coordination with the NGSO operators on prearranged DL modes, telescope observation time window, receiver frequency range, and certain sky regions that the telescope's main beam would be pointing towards. These tests typically consist of close communication between NRAO and the NGSO engineering teams at least two weeks ahead of the scheduled test time. Final details are typically confirmed within 24--48~hours before the telescope observation is conducted. The coordinated tests typically last for 0.5--2~hours to allow both parties to explore operational parameters and extreme scenarios. Once certain operational and coordination modes are established, NRAO can conduct longer telescope observation sessions, often without the need to coordinate with the NGSO operator. This type of follow-up observation serves as a blind experiment to ensure the coordination is performing as expected. For example, NRAO has conducted around several dozen coordinated tests with SpaceX's Starlink on the VLA, GBT, and VLBA since 2021. Many of these tests have resulted in publications, including reports \cite{depree2023memo154, depree2023memo222, depree2023memo223, bautista2025Lband, bautista2025uwbr, lawson2026alamo} and peer-reviewed journal articles \cite{nhan2024toward, nhan2025ods}; many of these have been coauthored by NRAO and SpaceX. 

\section{Spectrum Interference Avoidance Schemes}
\label{sec:avoidance}
Conventionally, passive spectrum services like RAS and Earth Exploration Satellite Services (EESS) would just flag and excise the corrupted data channels by RFI from scientific analysis. However, this is not sustainable as the amount of spectrum being reallocated for broader commercial utilization increases. It is important for both passive and active spectrum users to adopt new approaches to dynamically share the spectrum. In general, the sharing mechanism can be reactive or predictive, either using real-time spectrum sensing or a commonly shared database. 

Within the NRDZ communities, there have been other avoidance techniques being explored, such as the Satellite Orbit Prediction Processor (SOPP) predictive algorithm that both anticipates potential incoming satellite RFI and then optimizes the telescope scheduling to observe a relatively cleaner sky region \cite{gifford2024satellite}, the real-time geofencing (RTG) for protecting EESS radiometer measurements from 5G or 6G mm-wave transmissions \cite{eichen2024rf}, and the Adaptive Spectrum Tuning and Reactive Allocation (ASTRA) spectrum management framework for transmitter access \cite{sarbhai2024reactive}.

\begin{figure*}[htb!]
  \centering
  {\includegraphics[width=0.75\linewidth]{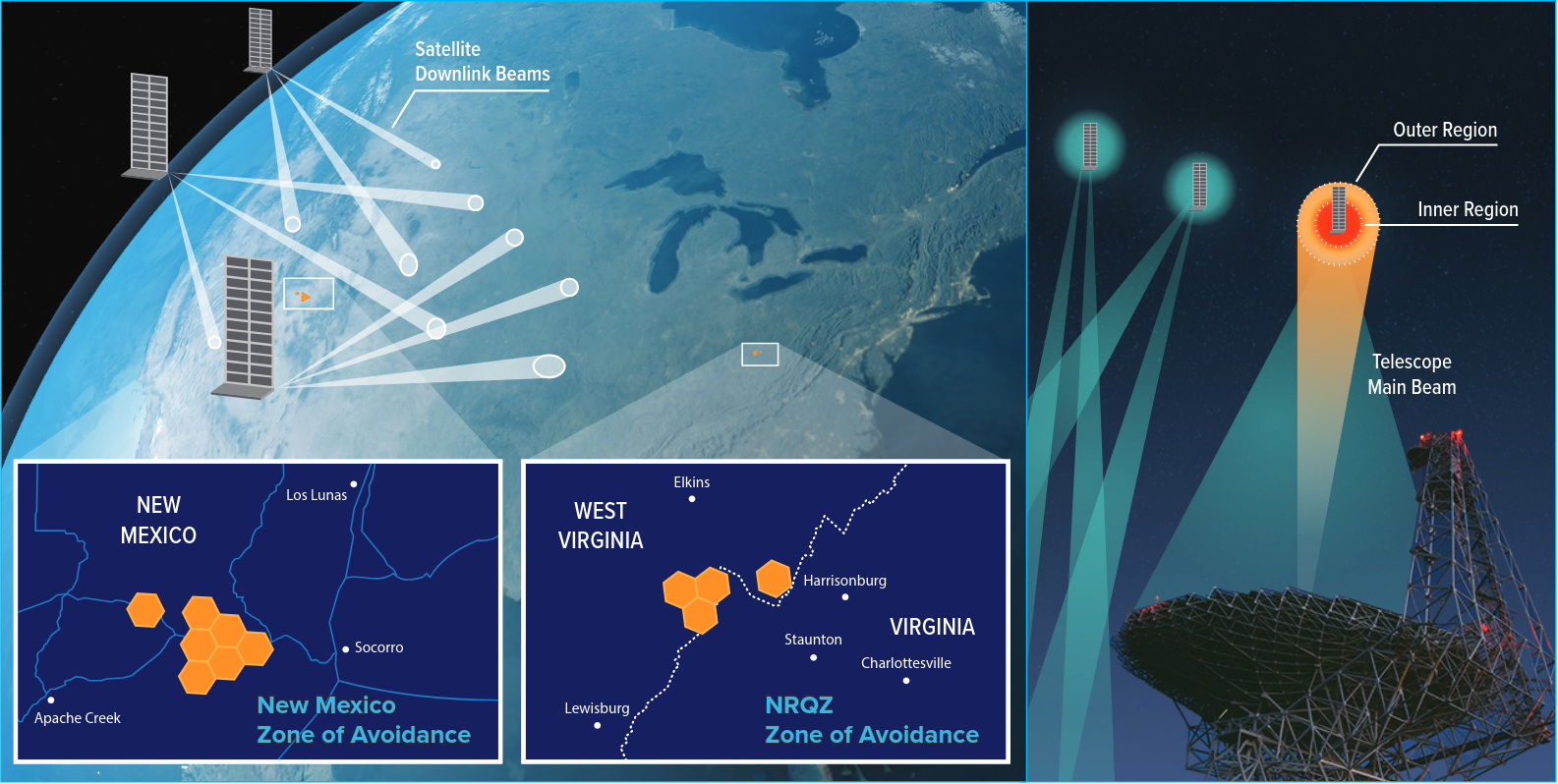}}
  \caption{A simplified illustration of the two approaches enabling spectrum coexistence between radio telescopes and satellite operators: (Left) Zone avoidance (ZA) around the VLA and GBT sites, (Right) Telescope boresight avoidance (TBA) with its inner and outer modes.}
  \label{fig:ods_illustration_new2025}
\end{figure*}

Previous coordinated experiments helped NRAO and other RAOs to devise several spectrum coexistence strategies with NGSO operators. They involve adopting avoidance techniques in three primary domains: geography, time, and frequency. These schemes are illustrated in Figure~\ref{fig:ods_illustration_new2025}. 

\subsection{Zone Avoidance (ZA)}
\label{sec:za}
Generally, if satellite operators can avoid directly illuminating geographical regions where radio telescopes are located at with the main lobe of their DL beams, RFI levels will be lower and confined within their allocated transmitting bands as expected. We refer to such a technique, analogous to the RTG, that SpaceX adopted for NRAO/GBO telescopes as Zone Avoidance (ZA). 
%RTG is defined? 

In principle, NGSO operators can adopt similar ZA by simply knowing the geographical coordinates of each of the radio telescopes, many of which are provided in the FCC Rule US131 \cite{fccus131} including areas for the GBO and the Sugar Grove Research Station (SGRS) within the National Radio Quiet Zones (NRQZ), as well as the VLA and VLBA's Pie Town (VLBA-PT) station in the New Mexico Radio Coordination Zone (NMRCZ). This coordination does not require any real-time data exchange with the telescopes. However, the projected area on the ground equivalent to this zone of avoidance (ZoA) is highly dependent of the main lobe size of the telescope and satellite's DL beams at a given frequency band of interest and orientation of the satellites approaching the telescope. Meanwhile, lower frequency transmissions typically result in larger beam widths and thus larger ZoA. Since each radio telescope and NGSO systems are unique, close coordination is needed when determining the ZoA parameters. 

As part of the pre-satellite deployment coordination, the ZA was crucial in mitigating large impacts to the radio telescope over the majority of observational time. In this stage, NRAO reviewed the technical information provided by an NGSO operator, most importantly the non-public downlink antenna beam pattern or mask, in respect to the ITU-R RA.769-2 or NRAO NRQZ power flux density threshold. A receiver sidelobe level of 0~dBi is typically assumed based on ITU recommendations for radio astronomy telescope sidelobes in interference studies. The end result was a setback distance from the telescope location within which the NGSO would not place their downlink’s main beams. During the process, NRAO noted this situation would not constitute a worst-case scenario due to the telescope's main beam gain not being accounted for (e.g., a satellite going through the telescope’s boresight), however it does provide a mitigation method for the majority of observational time. See Section~\ref{sec:tba} for additional mitigation techniques related to this worst-case scenario. Real world testing, additional technical parameters, and advanced modeling may shift this ZoA over time, thus warranting continuing collaboration between all stakeholders.

\subsection{Telescope Boresight Avoidance (TBA)}
\label{sec:tba}
A small number of satellite passages will occasionally traverse across sky regions where a telescope's main beam is pointing. This potential interaction increases the likelihood of the NGSO main beam interacting with the telescope's main beam. This will result in the strongest DL being measured by the telescopes if they happen to use a receiver band that overlaps with the DL frequencies. Such a strong interaction often causes the telescopes' front-end (FE) electronics, such as the sensitive custom-made low-noise amplifiers (LNAs), to saturate or compress, and results in significant drops in amplification. Frequent exposure of the LNAs to those close encounters can even potentially damage them. If the FE is not saturated, the strong DL signal can also lead to overflow in the signal digitization stage in the telescopes' analog-to-digital converters (ADCs), which results in spectral leakage across all frequency channels in the measured spectra at those timestamps, even if the NGSO DL signal is well confined within its allocated bands. 

To prevent these incidents in temporal and spectral domains, NRAO and SpaceX have developed the Telescope Boresight Avoidance (TBA) technique together. Previous coordinated tests demonstrate that RFI from these strong close encounters could be suppressed if SpaceX adapted Starlink's DL signal accordingly when the satellites cross the telescope's main beam (or boresight) \cite{nhan2024toward}. The TBA consists of two modes: 
\begin{enumerate}
    \item \textit{Outer TBA} - A satellite would only place its beams as far as possible from the radio telescope (typically 180~km away), which is then only illuminated by the DL beam's sidelobes, when it is getting close to the telescope main beam but outside the critical angular separation from the telescope boresight (typically less than 10--12$^{\circ}$, depending on the band),
    \item \textit{Inner TBA} - A satellite would momentarily disable beam forming and downlink when it enters a few degrees or less from the boresight.
\end{enumerate}

\subsection{Operational Data Sharing (ODS) System}
\label{sec:ods}
\begin{figure}[htb!]
  \centering
  {\includegraphics[width=1\linewidth]{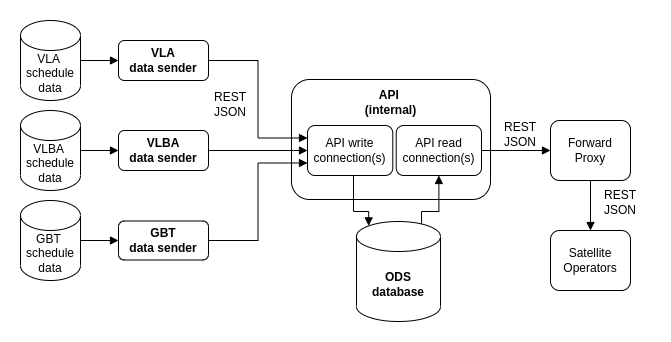}}
  \caption{A high level block diagram of the current NRAO ODS server showing how NRAO’s VLA, VLBA, and GBO’s GBT report their observation schedules as MRs to the ODS server for satellite operators using respective schedule Data Sender software at each telescope. In return, satellite API clients can post relevant auxiliary data (like TBA verification logs) back to the ODS server.}
  \label{fig:ods_api_block_diagram}
\end{figure}

To automate the information exchange between RAS and NGSO during the close encounters, NRAO has led the development of the Operational Data Sharing (ODS) system \cite{nhan2025ods}. The NRAO ODS is a machine-to-machine (M2M) service that reports upcoming NRAO/GBO telescope observation schedules to the ODS server from which NGSO operators can query the telescope information in near real-time using an Application Programming Interface (API) to task their satellites for enabling the TBA scheme (Figure~\ref{fig:ods_api_block_diagram}). The ODS API entries, or Mitigation Requests (MRs), are reported in JSON format and contain the key parameters of the upcoming telescope observation configuration, including astronomical source pointing coordinates (in Right Ascension and Declination), the time at which an observation starts and ends (in UTC), and the frequency coverage range of the chosen telescope receiver (in Hz), along with other relevant fields which are available on \url{https://obs.vla.nrao.edu/ods}. 

As of spring 2026, the NRAO ODS server is operational for all scripted VLA and VLBA observations, along with a small subset of GBT observations in development mode. At the moment, since the Starlink system only transmits in the 1990-1995~MHz DTC band and the 10.7-12.7~GHz for their broadband internet band, SpaceX only engages TBA for ODS MRs overlapping these two frequency ranges, in conjunction with the ZA for VLA, a subset of VLBA stations, and GBT. 

\begin{figure*}[htb!]
    %\centering
    \gridline{\includegraphics[width=0.9\linewidth]{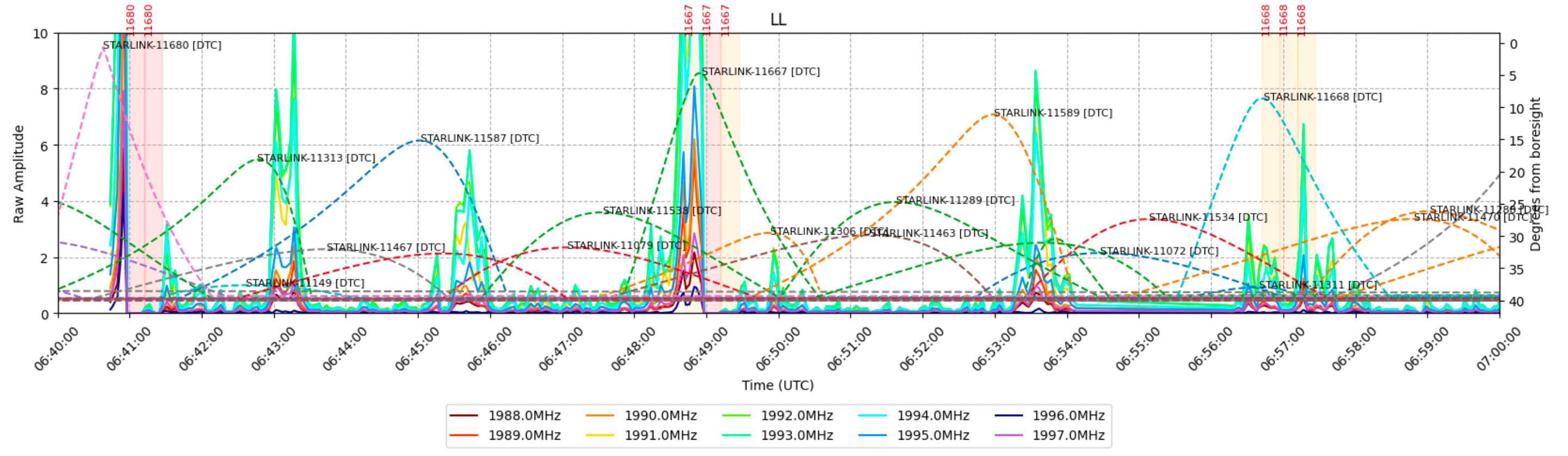}}%{(a)}
    \gridline{\includegraphics[width=0.9\linewidth]{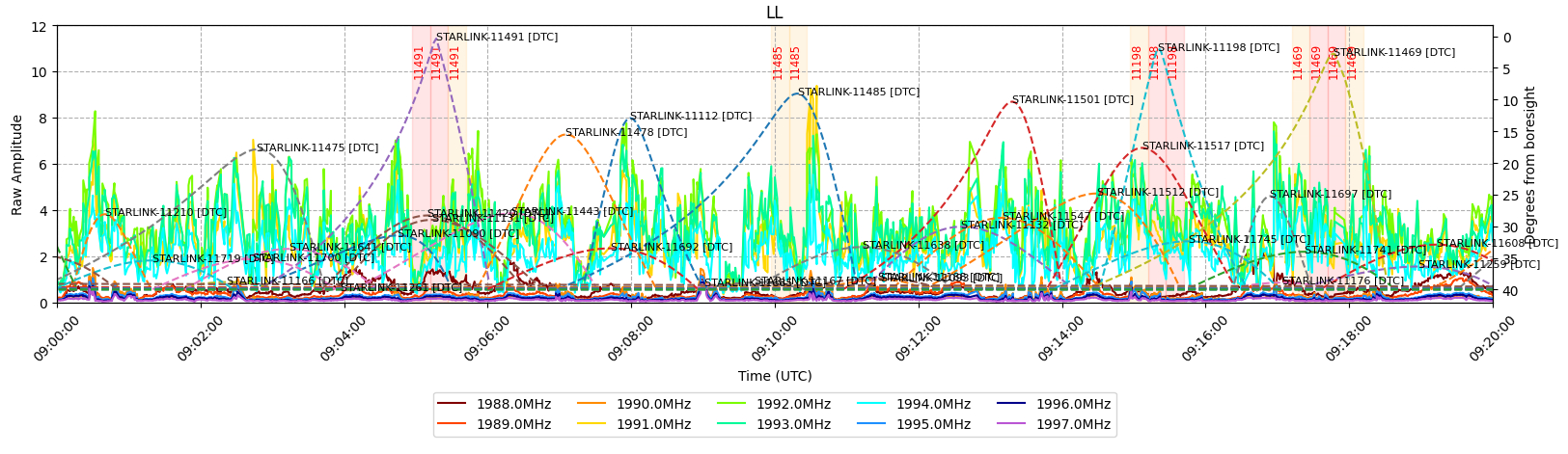}}%{(b)}
    \caption{Two examples illustrate TBA being activated by SpaceX in the DTC band (1990-1995~MHz) for the close-to-boresight encounters during observations at the VLA in October 2025 (top) and VLBA's Fort Davis station in January 2026 (bottom). The raw visibility amplitudes for the auto-correlated left-handed circular polarization of one of the antenna (solid curves) are plotted with the angular boresight separation (dashed curves) over approximately 20 minutes of observation for the expected Starlink satellite passages. The measured DL signal level drops as the satellite crosses the telescope boresight (i.e., boresight separation approaches $0^{\circ}$) when the outer (orange shaded) and inner (red shaded) TBA modes were engaged. Particularly, the DL signal vanishes as expected during the inner TBA when beam formation is briefly disabled. The TBA events last typically less than a minute at the DTC band, whereas they last only on the order of a few seconds at the broadband internet frequencies (10.7-12.7~GHz) due to smaller DL beams.}
    \label{fig:ods_spec_LL}
\end{figure*}

Since January 2025, as a means to transparently validate TBA events, SpaceX has been providing NRAO informative satellite avoidance logs to indicate the timestamps and satellite ID when the TBA is engaged by the given Starlink satellite. These logs are used by the NRAO team to validate the success of the TBA events by correlating each event with a decrease in downlink power received by the telescope. We also calculate the relative angular separation between a given satellite and the telescope boresight pointing when it is tracking an astronomical source across sky, using publicly available Two-Line Element (TLE) data for the satellites' ephemerides. The effectiveness of the SpaceX's TBA is illustrated in Figure~\ref{fig:ods_spec_LL} for recent VLA and VLBA observations, where the received amplitude as a function of time show a distinct decrease in signal as the Starlink satellites approach the telescope boresight due to the satellites redirecting and deactivating their downlink capabilities during the inner TBA.

\section{Case Study: ODS community participation}
\label{sec:case_study}
The initial SWIFT-SAT scope for ODS was meant to be used only by NRAO and SpaceX. However, as the ODS system and operational schemes have matured, NSF has encouraged wider participation in ODS by other stakeholders.

For RAS, other RAOs have adopted the same ODS JSON Schema for the critical fields and set up their own ODS API servers independently to interface with SpaceX for their own coordination. This includes the US-based HCRO and Haystack, which are currently running their ODS with TBA activated by SpaceX. The DSA-2000 and KPGO teams are conducting limited TBA testing with SpaceX. Additionally, the Commonwealth Scientific and Industrial Research Organisation's (CSIRO) Australia Telescope National Facility (ATNF) has also established its own ODS server to report their telescopes' data, including the Australia Telescope Compact Array (ATCA), Parkes Observatory, Mopra Observatory, Australian Square Kilometre Array Pathfinder (ASKAP), to use TBA with SpaceX.

In July 2025, NSF hosted the first Worldwide ODS Workshop (WOW) with virtual participation of US-based and international radio astronomy sites, along with geodesy communities, and SpaceX. This initial meeting provided an opportunity for initial information gathering on potential use cases and requirements from other radio telescope facilities if they were to adopt the ODS standards. 

For NGSO operators, with the additional coordination agreements NSF has signed in 2025, API access credentials have been provided to the Amazon Leo and AST SpaceMobile teams to interface with the NRAO ODS server for preliminary testing when their systems are more fully deployed. 

\section{Lessons Learned}
\label{sec:lessons}
Much of the progress made in ODS and TBA development in the past few years has been made possible by building mutual trust. The core of this trust is understanding and respect for each other's core missions. While the core missions of NRAO and Starlink are very different, both organizations have complex technical missions and highly engaged engineering and scientific staff. Cultivating an environment in which these groups can interact frequently, perform coordinated experiments, and engage in frequent discussions has been essential to the success of this work.

Some experiments are carried out by NRAO independently, and others have required coordinated work. NRAO has found that particularly for the coordinated experiments, being able to coauthor with NGSO operators encourages transparency and mutual understanding in the published results.

Another crucial aspect for the success of coordination is a mutual understanding in exchanging any raw data or sensitive information. As academia collaborates with commercial and federal parties, one needs to be mindful when requesting certain types of data from the collaborators. It helps to support the request tremendously if one can illustrate the use cases and usefulness of the shared data, as well as being clear how the data is being handled for storage, analysis, and publication. Many of these data management aspects can be incorporated in the coordination agreements. Furthermore, instead of asking for raw data, it may be worthwhile to explore the possibility of using reduced data that can easily serve the original purpose. For example, SpaceX has been able to share the TBA logs routinely for our telescope observation since they do not contain any sensitive information of their system and customers. The shared TBA logs contains only the essential information for NRAO to validate whether the TBA modes were activated at the expected times compared to the observational data, as shown in Figure~\ref{fig:ods_spec_LL}. 

Furthermore, not all radio telescopes and NGSO systems are identical. There is no "one size fits all" solution for all parties. One needs to explore different hardware, software, and operational policies to best fit each partner individually. For example, even adopting the same ODS JSON standard and general coordination procedures, it is necessary for each RAO to refine their testing settings and avoidance parameters with NGSO operators individually for their telescopes. This will hold true for future stakeholders who wish to participate in a similar coordination strategy. Importantly, due to the dynamic nature of all spectrum usage, it is imperative to constantly monitor and conduct periodic testing to ensure the operational continuity of the adopted coordination strategy.

\section{Path forward and possible future plans}
\label{sec:outlooks}
By design, the ODS is frequency agnostic and has been reporting all receiver frequency ranges whenever they are being scheduled on the telescopes. This allows the ODS system to be adopted by other NGSO systems transmitting at different bands.

The ODS API system does not need to be limited to just coordination between RAS and NGSO. In essence, ODS provides the API and database for information exchange between different spectrum services (both passive and active) that may be able to use it for coordination; these uses include, but are not limited to: (i) mutual awareness of each other’s operational and frequency band occupancy status, (ii) facilitation of avoidance schemes to achieve spectrum coexistence and minimize mutual interference.

Depending on the level of coordination and participation, some of the possible future ODS use cases may include other spectrum services, such as terrestrial transmission, EESS and commercial satellites with their Earth stations. In fact, NSF and NRAO have recently started some preliminary coordinated testing with a cellular provider to explore means to mitigate RFI generated by a newly upgraded tower near the VLA. These efforts involve manual coordination between the cellular engineering team and NRAO spectrum management team, running propagation simulations, and conducting test observations on the telescope to evaluate the RFI levels generated by different operational modes of the cellular tower. Some aspects of this process could potentially be automated by utilization of a service like ODS.

\begin{figure}[htb!]
  \centering
  {\includegraphics[width=1\linewidth]{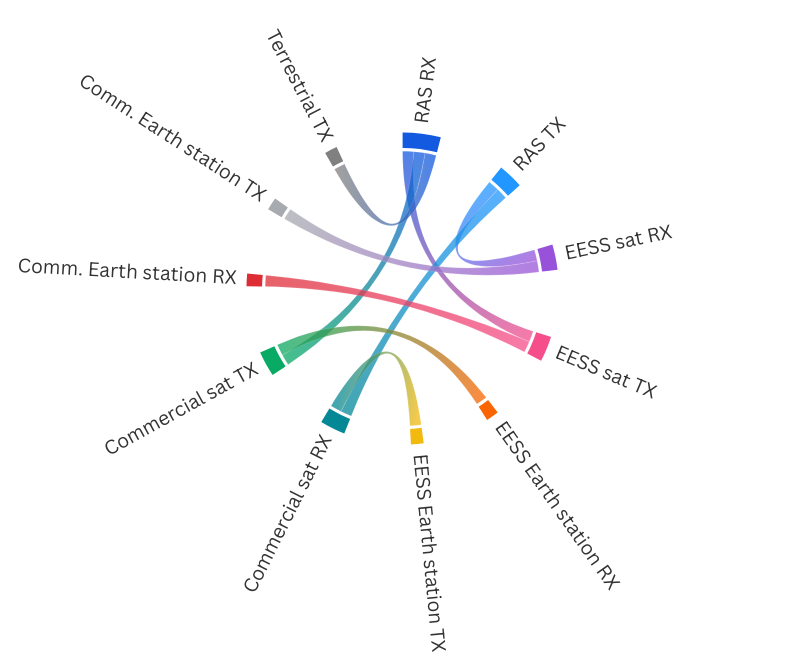}}
  \caption{A chord diagram illustrates potential future use cases of the ODS system for interchanging spectrum coordination data between different services, including both active and passive spectrum users for their transmitting (TX) and receiving (RX) operations. This eventual expansion will require a joint development of data standards and close coordination in testing and validation among all participants.}
  \label{fig:ods_use_case_chord_diagram}
\end{figure}

These future stakeholders can potentially interchange information with the ODS server as illustrated in Figure~\ref{fig:ods_use_case_chord_diagram}, which schematically diagrams a wide variety of possible interactions. It will be essential to have a joint development and mutual understanding on protocols and software standards among all API clients. Nonetheless, the NRAO development team is actively working on scaling the ODS server infrastructure up to ensure service reliability when accommodating additional external clients in the next few years. 

\iffalse
between receiving (RX) and transmitting (TX) services may include, but not limited to, 
\begin{itemize}
    \item RAS RX vs. Terrestrial cellular TX,
    \item RAS RX vs. Terrestrial federal TX,
    \item RAS RX vs. EESS sat. downlink,
    \item RAS TX vs. commercial/EESS sat. RX,
    \item EESS Earth station RX vs. commercial sat. TX,
    \item Commercial sat. Earth station RX vs. EESS sat. TX,
\end{itemize}
where RAS may involve radar transmission, e.g., ngRadar \cite{armentrout2026ngradar}.
\fi

\section*{Acknowledgments}
The NRAO and GBO are facilities of the U.S. NSF operated under cooperative agreement by Associated Universities, Inc (AUI). This work is supported by the NSF's SII-NRDZ (AST-2232159) and SWIFT-SAT (AST-2332422) grants. The current NRAO ODS core development team consists of: Dawn Pattison, Daniel Faes, Thomas Chamberlin, Bang Nhan, Mark Whitehead, and Randall Arnold. The authors acknowledge the contributions of many individuals from NRAO/GBO and NGSO operators who have made these coordination possible. Particularly, the authors thank the SpaceX engineering team, including Christopher Steele, Matt Iverson, Jacob Donenfeld, and Daniel Dueri. The authors also acknowledge discussion and collaboration with the Amazon Leo (formerly Kuiper) and AST SpaceMobile teams. The authors would also like to acknowledge our collaborators for adopting and testing the ODS standards independently for their respective telescope systems:  David DeBoer (UC Berkeley), Philip J. Erickson (MIT Haystack), Greg Hellbourg (Caltech OVRO), Balthasar Indermuehle (CSIRO), Alexander Pollak (SETI), Chris Coughlin (KPGO), and Phillip Haftings (USNO).

%{\appendices
%\section*{Proof of the First Zonklar Equation}
%Appendix one text goes here.
% You can choose not to have a title for an appendix if you want by leaving the argument blank
%\section*{Proof of the Second Zonklar Equation}
%Appendix two text goes here.}

\bibliographystyle{IEEEtran}
\bibliography{IEEEabrv,reference_ods} 

\end{document}